\documentclass[preprint,12pt]{elsarticle}
\journal{Computer Physics Communications}
\usepackage[T1]{fontenc}
\usepackage{framed}
\usepackage[utf8]{inputenc}
\usepackage{url}
\usepackage{graphicx}
\usepackage{amsmath}
\usepackage{amssymb}
\usepackage{hyperref}
\usepackage{longtable}
\usepackage{color}
\graphicspath{{figures/}{fig/}}

\newcommand{\ui}{\mathrm{i}}
\newcommand{\ud}{\mathrm{d}}

\begin{document}
\begin{frontmatter}

\title{{\tt AMFlow}: a Mathematica package for Feynman integrals computation via Auxiliary Mass Flow}

\author[a,b]{Xiao Liu\corref{author}}
\ead{xiao.liu@physics.ox.ac.uk}

\author[a,c,d]{Yan-Qing Ma}
\ead{yqma@pku.edu.cn}

\cortext[author] {Corresponding author.}

\address[a]{School of Physics and State Key Laboratory of Nuclear Physics and Technology, Peking University, Beijing 100871, China}
\address[b]{Rudolf Peierls Centre for Theoretical Physics, Clarendon Laboratory, Parks Road, Oxford OX1 3PU, UK}
\address[c]{Center for High Energy Physics, Peking University, Beijing 100871, China}
\address[d]{Collaborative Innovation Center of Quantum Matter, Beijing 100871, China}

\begin{abstract}
{\tt AMFlow} is a Mathematica package to numerically compute dimensionally regularized Feynman integrals via the recently proposed auxiliary mass flow method. In this framework, integrals are treated as functions of an auxiliary mass parameter and their results can be obtained by constructing and solving linear differential systems with respect to this parameter, in an automatic way. The usage of this package is described in detail through an explicit example of double-box family involved in two-loop $t\bar{t}$ hadroproduction.
\end{abstract}


\end{frontmatter}

\newpage

\textbf{PROGRAM SUMMARY}

\vspace{1cm}

\begin{small}
\noindent
{\em Program title:} {\tt AMFlow}\\
{\em Developer's repository link:} \url{https://gitlab.com/multiloop-pku/amflow}\\
{\em Licensing provisions:} MIT\\
{\em Programming language:} {\tt Wolfram Mathematica} 11.3 or higher\\
{\em External routines/libraries used:} {\tt Wolfram Mathematica} [1], {\tt FiniteFlow} [2], {\tt LiteRed} [3], {\tt Kira} [4], {\tt FIRE} [5]\\
{\em Nature of problem:}
Automatically obtaining high-precision numerical results for dimensionally regularized Feynman integrals at arbitrary points in phase-space. \\
{\em Solution method:}
The program implements recently proposed auxiliary mass flow method, which introduces an auxiliary mass parameter to Feynman integrals and solves differential equations with respect to this parameter to obtain physical results. \\
{\em Restrictions:} the CPU time and the available RAM \\
{\em References:}
{\\} [1] \url{http://www.wolfram.com/mathematica}, commercial algebraic software;
{\\} [2] \url{https://github.com/peraro/finiteflow}, open source;
{\\} [3] \url{http://www.inp.nsk.su/~lee/programs/LiteRed}, open source;
{\\} [4] \url{https://gitlab.com/kira-pyred/kira}, open source;
{\\} [5] \url{https://bitbucket.org/feynmanIntegrals/fire}, open source.
\end{small}

\newpage

\section{Introduction}\label{sec:intro}

Computation of Feynman integrals is crucial for the purpose of testing the standard model of particle physics and probing new physics. Currently, the main strategy is to first reduce all Feynman integrals in a problem to a small set of bases~\cite{Chetyrkin:1981qh, Laporta:2000dsw, Gluza:2010ws, Schabinger:2011dz, vonManteuffel:2012np, Lee:2013mka, vonManteuffel:2014ixa, Larsen:2015ped, Peraro:2016wsq, Mastrolia:2018uzb, Liu:2018dmc,  Guan:2019bcx, Klappert:2019emp, Peraro:2019svx, Frellesvig:2019kgj, Wang:2019mnn, Smirnov:2019qkx, Klappert:2020nbg, Boehm:2020ijp, Heller:2021qkz, Bendle:2021ueg}, called master integrals, and then calculate these master integrals.

There are many methods on the market to compute master integrals, such as: sector decomposition~\cite{Hepp:1966eg, Roth:1996pd, Binoth:2000ps, Heinrich:2008si, Smirnov:2015mct, Borowka:2015mxa, Borowka:2017idc}; Mellin-Barnes representation~\cite{Boos:1990rg, Smirnov:1999gc, Tausk:1999vh, Czakon:2005rk, Smirnov:2009up, Gluza:2007rt}; difference equations~\cite{Laporta:2000dsw, Lee:2009dh}; traditional differential equations~\cite{Kotikov:1990kg, Kotikov:1991pm, Remiddi:1997ny, Gehrmann:1999as, Argeri:2007up, MullerStach:2012mp, Henn:2013pwa, Henn:2014qga, Moriello:2019yhu, Hidding:2020ytt}, by setting up and solving differential equations satisfied by master integrals with respect to kinematic variables $\vec{s}$;
and others \cite{Catani:2008xa,Rodrigo:2008fp,Bierenbaum:2010cy,Bierenbaum:2012th,Tomboulis:2017rvd,Runkel:2019yrs,Capatti:2019ypt,Aguilera-Verdugo:2020set,Song:2021vru,Dubovyk:2022frj}. The sector decomposition method and Mellin-Barnes representation method can be applied in principle to any integral. However, it is well known that these methods, which need to calculate multidimensional integrations directly, are very inefficient to obtain high-precision results. Difference equations and differential equations can be very efficient, but they depend on integrals reduction to set up relevant equations, which may become very nontrivial for multiloop multiscale problems. Besides, usually there is no systematic way to obtain boundary conditions for these two methods.

The auxiliary mass flow method~\cite{Liu:2017jxz,Liu:2021wks} is also a kind of differential equations method, which calculates Feynman integrals by setting up and solving differential equations with respect to an auxiliary mass term $\eta$. This method has many advantages. First, it is systematic, because boundary conditions at $\eta\to\infty$ can be obtained iteratively\cite{Liu:2021wks,Liu:2022mfb}.  Second, as only ordinary differential equations are involved, high-precision results can be efficiently obtained  \cite{Wason:1987aa}. Third, integrals containing linear propagators and phase-space integrations can all be calculated \cite{Liu:2022tji, Liu:2020kpc}. Finally, integrals reduction to set up differential equations with respective to $\eta$ is usually easier than to set up differential equations with respective to $\vec{s}$~\cite{Liu:2021wks}.
Therefore, as long as reduction tools are powerful enough to set up differential equations with respect to $\eta$, auxiliary mass flow can always provide high-precision result efficiently.

Auxiliary mass flow method has already been used to study many physical processes \cite{Zhang:2018mlo,Yang:2020msy,Bronnum-Hansen:2020mzk,Zhang:2020atv,Bronnum-Hansen:2021olh,Wu:2021tzo,Bronnum-Hansen:2021pqc,Baranowski:2021gxe, Sang:2022kub, Chaubey:2022hlr, Zhang:2022jqw,Abreu:2022vei, Feng:2022vvk, Feng:2022ruy, Sang:2022erv}. Especially, equipped with the iterative strategy~\cite{Liu:2021wks}, this method becomes extremely powerful so that many Feynman integrals in cutting-edge problems, which are very challenging for other methods, can be calculated (see Fig. 4 in Ref.~\cite{Liu:2021wks}). It is thus valuable for high-precision phenomenological studies.
This paper aims to provide a public implementation of this method, including the automation of the fully iterative strategy and a high-performance numerical solver for ordinary differential equations, so that it can be more widely used for phenomenological studies. 

\section{Auxiliary mass flow}\label{sec:amflow}
In this section we give a review of the auxiliary mass flow method, concentrating on the computation of normal loop integrals~\cite{Liu:2017jxz, Liu:2021wks}. The extensions to compute integrals containing linear propagators or phase-space integrations can be found in Refs.~\cite{Liu:2022tji, Liu:2020kpc}.

\subsection{The plain method}
Let us consider a dimensionally regularized Feynman integral family defined by
\begin{align}\label{eq:ampara0}
I(\vec{\nu}, \vec{s}, \epsilon)=\int\prod_{i=1}^{L}\frac{\ud^{D}\ell_i}{\ui\pi^{D/2}}
\frac{\mathcal{D}_{K+1}^{-\nu_{K+1}}\cdots \mathcal{D}_N^{-\nu_N}}{(\mathcal{D}_1+ \ui 0^+)^{\nu_{1}}\cdots (\mathcal{D}_K+\ui 0^+)^{\nu_{K}}},
\end{align}
where $\vec{s}$ is the list of all kinematic variables including Mandelstam variables and nonzero masses of particles, $D = 4-2\epsilon$ is the spacetime dimension, $L$ is the number of loops, $\ell_i$ are loop momenta, $\mathcal{D}_1,\ldots,\mathcal{D}_K$ are inverse propagators, $\mathcal{D}_{K+1},\ldots,\mathcal{D}_N$ are irreducible scalar products introduced for completeness,  $\nu_1,\ldots,\nu_K$ can be any integers, and $\nu_{K+1},\ldots,\nu_N$ can only be non-positive integers. We next introduce an auxiliary integral family by inserting an auxiliary parameter $\eta$ to each propagator of \eqref{eq:ampara0}
\begin{align}\label{eq:ampara}
I_\text{aux}(\vec{\nu}, \vec{s}, \epsilon, \eta)=\int\prod_{i=1}^{L}\frac{\ud^{D}\ell_i}{\ui\pi^{D/2}}
\frac{\mathcal{D}_{K+1}^{-\nu_{K+1}}\cdots \mathcal{D}_N^{-\nu_N}}{(\mathcal{D}_1-\eta)^{\nu_{1}}\cdots (\mathcal{D}_K- \eta)^{\nu_{K}}}.
\end{align}
Then physical results can be recovered by taking the following limit
\begin{align}\label{eq:def}
I(\vec{\nu},\vec{s},\epsilon)= \lim_{\eta\to \ui 0^-}{I}_\text{aux}({\vec{\nu}} , \vec{s}, \epsilon, \eta).
\end{align}

This auxiliary family, although seems to be more complicated than the original one, becomes rather simple as $\eta$ approaches the infinity. This can be understood through region analysis~\cite{Beneke:1997zp, Smirnov:1999bza}. More specifically, when $|\eta|$ is very large, only the integration region with $\ell_i^\mu \sim \mathcal{O}(\sqrt{\eta})$ can contribute, and thus every propagator can be expanded like
\begin{align}\label{eq:deexpand}
\frac{1}{((\ell+p)^2 -m^2 -\eta)^\nu} = \frac{1}{(\ell^2-\eta)^\nu} \sum_{i = 0}^\infty \frac{(\nu)_i}{i!} \left(-\frac{2\ell\cdot p+p^2-m^2}{\ell^2-\eta}\right)^i,
\end{align}
where $(\nu)_i\equiv\Gamma(\nu+i)/\Gamma(\nu)$ is the Pochhammer symbol. After all such kinds of expansion, what we get are combinations of equal-mass vacuum integrals, which have been intensively studied in literature~\cite{Davydychev:1992mt, Broadhurst:1998rz, Schroder:2005va,Luthe:2015ngq, Luthe:2017ttc, Kniehl:2017ikj}. As a result, auxiliary integrals $I_\text{aux}(\vec{\nu}, \vec{s}, \epsilon, \eta)$ in the neighborhood of $\eta = \infty$ can be easily obtained and what remains is to perform analytic continuation (\textit{auxiliary mass flow}) of them to recover physical results.

As auxiliary integrals can be expressed as linear combination of master integrals using integrals reduction, we only need to perform analytic continuation for master integrals, denoted by the vector $\vec{\mathcal{I}}_\text{aux}(\vec{s},\epsilon, \eta)$.
Integrals reduction can also setup differential equations for master integrals, which look like
\begin{align}\label{eq:deq}
	\frac{\partial}{\partial \eta}\vec{\mathcal{I}}_\text{aux}(\vec{s},\epsilon, \eta) = A(\epsilon, \eta)\vec{\mathcal{I}}_\text{aux}(\vec{s},\epsilon, \eta).
\end{align}
For any fixed generic kinematic configuration $\vec{s} = \vec{s}_0$ \footnote{If possible, $\vec{s}_0$ should be chosen as some simple rational numbers (e.g., if one only wants to obtain boundary conditions for a system of differential equations). Because the computational cost of construction of differential equations, which is usually the dominant part for complicated cutting-edge problems, depends heavily on the choice of $\vec{s}_0$.}, the above differential equations can be numerically solved by using series expansions, similar to numerically solving differential equations with respective to kinematic variables~\cite{Caffo:2008aw, Czakon:2008zk}, which can realize the flow of $\eta$ from the boundary at $\infty$ to physical value at $\ui 0^-$.

Before describing how to solve the above differential equations, it is helpful to know some basic features of these auxiliary integrals as analytic functions of $\eta$. According to Cutkosky rules~\cite{Cutkosky:1960sp}, integrals can be only real-valued on the real axis when $\eta > \eta_\text{th}$, where $\eta_\text{th}$ is the largest threshold for the corresponding process.
Thus the branch cut of the auxiliary integral can be defined as the straight line connecting $\eta = -\infty$ and $\eta = \eta_\text{th}$ along the real axis, such that the Schwarz reflection principle
\begin{align}\label{eq:schwarz}
I_\text{aux}(\vec{\nu}, \vec{s}, \epsilon, \eta^*) = I_\text{aux}^*(\vec{\nu}, \vec{s}, \epsilon, \eta)
\end{align}
holds everywhere except the branch cut (for real $\vec{s}$ and $\epsilon$).

\begin{figure}[htbp]
\centering
\includegraphics[width=0.9\textwidth]{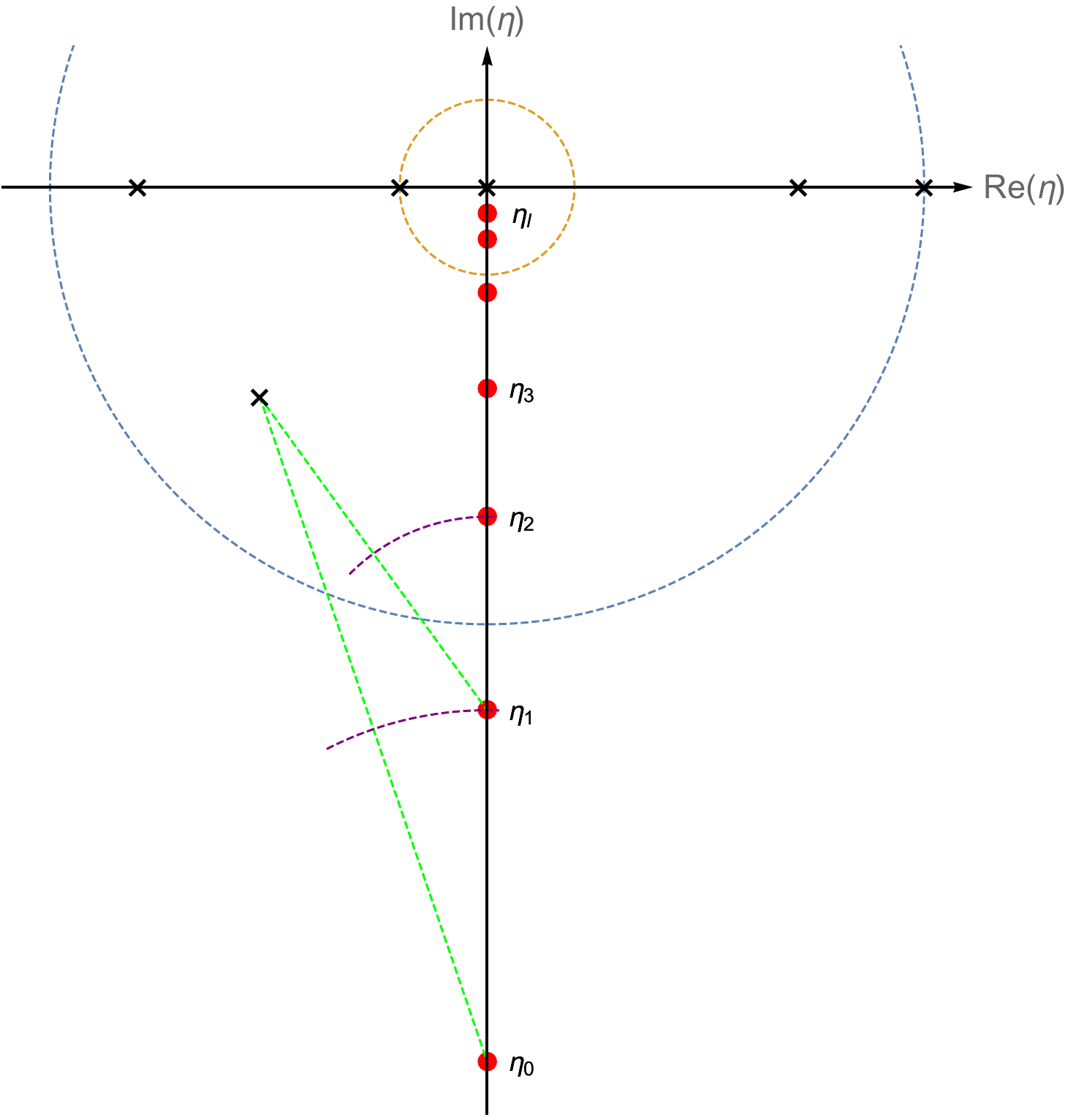}\\
\caption{Singularities and analytic continuations. Singularities are labeled as crosses. Solid dots are points where to perform series expansions.}\label{fig:pole}
\end{figure}

Now we can describe our strategy for analytic continuations, or solving differential equations.
We first need to define a path for the analytic continuations connecting $\eta = \infty$ and $\eta = \ui 0^-$, characterized by a list of regular points $\{\eta_0, \eta_1, \ldots, \eta_l\}$ on which we will perform series expansions in order. A typical choice is shown in Fig.~\ref{fig:pole}, where the larger (smaller) circle is defined as smallest (largest) circle centered at $\eta = 0$ that contains all singularities (no singularity) except $\eta = \infty (\eta = 0)$. The choice of the regular points should satisfy the following rules: i) $\eta_0$ is outside of the larger circle; ii) $\eta_l$ is inside the smaller circle; iii) the distance between $\eta_{i+1}$ and $\eta_i$ is smaller than the convergence radius of the series expansions centered at $\eta_i$.

Then the flow of auxiliary mass can be divided into three main stages: i) expanding the integrals around $\eta = \infty$ and estimating at $\eta = \eta_0$; ii) expanding at $\eta = \eta_i$ and estimating at $\eta = \eta_{i+1}$ for $i = 0,\ldots,l-1$; iii) expanding formally at $\eta = 0$ and matching at $\eta = \eta_l$ to determine the unknown coefficients in the formal asymptotic series. After these steps, we are able to take the limit $\eta \to \ui0^-$ for the expansion at $\eta = 0$ to obtain physical results.

A simple example would be helpful to explain the basic ideas of performing the expansions. For more technical details, see e.g. Refs.~\cite{Liu:2017jxz, Wason:1987aa}. Let us consider a massless one-loop two-point integral family
\begin{align}\label{eq:bubble}
I(\nu_1, \nu_2, \epsilon) = \left.\int\frac{\ud^D\ell}{\ui\pi^{D/2}}\frac{1}{(\ell^2+\ui0)^{\nu_1}((\ell+p)^2+\ui0)^{\nu_2}}\right|_{p^2=1}.
\end{align}
There is one master integral $I(1,1, \epsilon)$, whose result is
\begin{align}\label{eq:bubbleres}
I(1,1,\epsilon) &= (-1-\ui0)^{-\epsilon}\times\frac{\Gamma(\epsilon)\Gamma(1-\epsilon)^2}{\Gamma(2-\epsilon)}\nonumber\\
&=\frac{1}{\epsilon} + 1.42278+3.14159\ui+\mathcal{O}(\epsilon).
\end{align}
In the aforementioned auxiliary mass flow method, we first introduce the auxiliary mass parameter to obtain
\begin{align}\label{eq:bubbleaux}
I_\text{aux}(\nu_1, \nu_2, \epsilon, \eta) = \left. \int \frac{\ud^D\ell}{\ui\pi^{D/2}} \frac{1}{(\ell^2{-\eta})^{\nu_1} ((\ell+p)^2{ -\eta})^{\nu_2}}\right|_{p^2=1}.
\end{align}
Now there are two master integrals,
\begin{align}
\vec{\mathcal{I}}_\text{aux}(\epsilon,\eta) = \{I_\text{aux}(1,0, \epsilon, \eta), I_\text{aux}(1,1, \epsilon, \eta)\},
\end{align}
and the differential equations for them are
\begin{align}\label{eq:bubblede}
\frac{\partial}{\partial\eta}\vec{\mathcal{I}}_\text{aux}(\epsilon, \eta) = \left(
\begin{array}{cc}
 \frac{1-\epsilon }{\eta } & 0 \\
 \frac{2 (\epsilon -1)}{\eta  (4 \eta -1)} & -\frac{2 (2 \epsilon -1)}{4 \eta -1} \\
\end{array}
\right)\vec{\mathcal{I}}_\text{aux}(\epsilon, \eta).
\end{align}
Boundary condition for the first master integral can be computed fully analytically
\begin{align}\label{eq:bubblebc1}
I_\text{aux}(1,0,\epsilon,\eta) = \eta^{1-\epsilon}\times(-\Gamma(\epsilon-1)),
\end{align}
and the second one can be expanded near $\eta = \infty$ giving
\begin{align}\label{eq:bubblebc2}
I_\text{aux}(1,1, \epsilon, \eta) &=  \int \frac{\ud^D\ell}{\ui\pi^{D/2}} \frac{1}{(\ell^2{-\eta})^{2}} + \cdots\nonumber\\
&=\eta^{-\epsilon}\times(\Gamma(\epsilon)+\mathcal{O}(\eta^{-1})).
\end{align}
Next we define the list of regular points to perform expansions. We can read directly from the differential equations \eqref{eq:bubblede} that the singularities are $0$, $1/4$ and $\infty$. As a result, the list of regular points can be chosen as $\eta_0 = -\ui/2$, $\eta_1 = -\ui/4$ and $\eta_2 = -\ui/8$.

As the first master integral in this example has been totally solved, we just consider the second one. Near $\eta = \infty$, this integral can be expanded like
\begin{align}\label{eq:ansatzinf}
I_\text{aux}(1,1,\epsilon,\eta) = \eta^{-\epsilon} \sum_{n=0}^\infty a_n(\epsilon) \eta^{-n},
\end{align}
which is a natural generalization of its boundary condition \eqref{eq:bubblebc2}. We then substitute the expansions \eqref{eq:bubblebc1} and \eqref{eq:ansatzinf} into the differential equations \eqref{eq:bubblede} and what comes out is a system of recurrence relations which can be used to express $a_n(\epsilon)$ in terms of $a_0(\epsilon)$, the boundary input determined by Eq.\eqref{eq:bubblebc2}.
Some of the results are given in the following
\begin{align}
a_0(\epsilon) &= \frac{1}{\epsilon} - 0.577216+\mathcal{O}(\epsilon),\nonumber\\
a_1(\epsilon) &= 0.166667+\mathcal{O}(\epsilon),\nonumber\\
&\cdots\nonumber\\
a_{100}(\epsilon) &= 5.49443\times 10^{-64}+\mathcal{O}(\epsilon),
\end{align}
all of which are real-valued, if $\epsilon$ is real.
The expansion \eqref{eq:ansatzinf} enables us to estimate the value of $I_\text{aux}(1,1,\epsilon,\eta_0)$ through
\begin{align}\label{eq:valueateta0}
I_\text{aux}(1,1,\epsilon,\eta_0) \approx \eta_0^{-\epsilon}\sum_{n=0}^{100} a_n(\epsilon) \eta_0^{-n} = \frac{1}{\epsilon}+0.0548501+1.88709\ui+\mathcal{O}(\epsilon),
\end{align}
where
\begin{align}
\left(-\frac{\ui}{2}\right)^{-\epsilon}=1+\left(\frac{\ui\pi}{2}+\log(2)\right)\epsilon+\mathcal{O}(\epsilon^2)
\end{align}
has been used.
Expansion near the regular point $\eta = \eta_0$ is a Taylor expansion, which looks like
\begin{align}\label{eq:ansatzreg}
I_\text{aux}(1,1,\epsilon,\eta) = \sum_{n=0}^\infty b_n(\epsilon) (\eta-\eta_0)^n.
\end{align}
We again substitute this expansion along with the value of the first master integral into the differential equations \eqref{eq:bubblede} and obtain a system of recurrence relations, which can be used to reduce $b_n(\epsilon)$ to $b_0(\epsilon)$, the value of $I_\text{aux}(1,1,\epsilon,\eta_0)$ obtained in Eq.\eqref{eq:valueateta0}. Partial results of $b_n(\epsilon)$ are shown below
\begin{align}
b_0(\epsilon) &= \frac{1}{\epsilon}+0.0548501+1.88709\ui+\mathcal{O}(\epsilon), \nonumber\\
b_1(\epsilon) &= 0.5714-1.77538\ui+\mathcal{O}(\epsilon),\nonumber\\
&\cdots,\nonumber\\
b_{100}(\epsilon) &= -1.29958\times 10^{24}+1.28029\times 10^{26}\ui+\mathcal{O}(\epsilon).
\end{align}
Then we can estimate $I_\text{aux}(1,1,\epsilon,\eta_1)$ using the expansion near $\eta = \eta_0$ \eqref{eq:ansatzreg}
\begin{align}
I_\text{aux}(1,1,\epsilon,\eta_1) = \frac{1}{\epsilon}+0.609168+2.13174\ui+\mathcal{O}(\epsilon).
\end{align}
Similarly, we can expand near $\eta = \eta_1$ and obtain the estimation at $\eta=\eta_2$
\begin{align}\label{eq:bubblemat}
I_\text{aux}(1,1,\epsilon,\eta_2) = \frac{1}{\epsilon}+0.994236+2.42639\ui+\mathcal{O}(\epsilon).
\end{align}
At the last step, we need to consider the expansion near $\eta = 0$ and match at $\eta = \eta_2$. The general form of this expansion is
\begin{align}\label{eq:ansatzzero}
I_\text{aux}(1,1,\epsilon,\eta) = \sum_{n=0}^\infty c_n(\epsilon) \eta^{n} + \eta^{1-\epsilon}\sum_{n=0}^\infty d_n(\epsilon) \eta^{n},
\end{align}
where the left part comes from the homogeneous equation and the right part comes from the inhomogeneous equation (sub-topology). By substituting the expansions \eqref{eq:bubblebc1} and \eqref{eq:ansatzzero} into the differential equations \eqref{eq:bubblede}, we can obtain two sets of recurrence relations, which can be used to reduce all $c_n(\epsilon)$ to $c_0(\epsilon)$ and determine all $d_n(\epsilon)$ respectively. For example, we have
\begin{align}\label{eq:reccn}
c_1(\epsilon) &= 2(2\epsilon-1)c_0(\epsilon),\nonumber\\
c_2(\epsilon) &= 2(2\epsilon-1)(2\epsilon+1)c_0(\epsilon),\nonumber\\
&\cdots,
\end{align}
and
\begin{align}\label{eq:recdn}
d_0(\epsilon) &= -2\Gamma(\epsilon-1),\nonumber\\
d_1(\epsilon) &= 4\Gamma(\epsilon-1)/(\epsilon-2),\nonumber\\
&\cdots.
\end{align}
We find there is actually only one unknown parameter, $c_0(\epsilon)$, which can be determined through matching at $\eta=\eta_2$. By substituting the estimation at $\eta=\eta_2$ \eqref{eq:bubblemat} and the coefficients \eqref{eq:reccn} and \eqref{eq:recdn} into the series expansion \eqref{eq:ansatzzero}, we can  solve the resulting linear equation to obtain
\begin{align}
c_0(\epsilon) = \frac{1}{\epsilon} + 1.42278+3.14159\ui +\mathcal{O}(\epsilon).
\end{align}
After computing these expansions, we can finally take the physical limit $\eta \to \ui0^-$ in the expansion near $\eta = 0$ \eqref{eq:ansatzzero}. Note that in dimensional regularization, we have
\begin{align}
\lim_{\eta\to\ui0^-}\eta^{a+b\epsilon} = 0,
\end{align}
for any nonzero $b$. So what remains in this limit is just the leading term of the Taylor part, $c_0(\epsilon)$, i.e.,
\begin{align}
I(1,1,\epsilon) &\equiv \lim_{\eta\to\ui0^-}I_\text{aux}(1,1,\epsilon,\eta) \nonumber\\
&= c_0(\epsilon) \nonumber\\
&= \frac{1}{\epsilon} + 1.42278+3.14159\ui +\mathcal{O}(\epsilon),
\end{align}
which agrees with the analytic result \eqref{eq:bubbleres}.

\subsection{Iterative strategy}\label{sec:iterative}
One interesting phenomenon from the previous example is that the number of master integrals increases after introducing $\eta$. As a result, it can be expected for much more complicated problems, the introduction of $\eta$ may greatly increase the number of MIs, such that the differential equations \eqref{eq:deq} cannot be set up in reasonable time with current reduction techniques. To overcome this difficulty, in~\cite{Liu:2021wks} we propose to apply the auxiliary mass flow method iteratively to reduce the number of master integrals, and thus the computational cost, to a reasonable level.

\begin{figure}[htbp]
\centering
\includegraphics[width=0.4\textwidth]{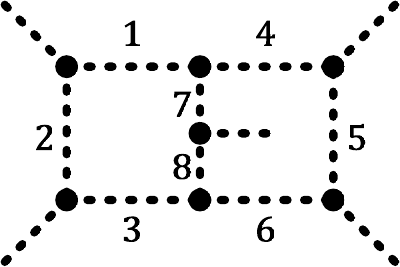}\\
\caption{A two-loop five-point massless double-pentagon topology.}\label{fig:dp}
\end{figure}

The key observation is that the number of master integrals can be reduced if $\eta$ is introduced to fewer propagators. For example, for the two-loop five-point massless double-pentagon integral family with 108 master integrals shown in \ref{fig:dp}, we obtain 476 master integrals if $\eta$ is introduced to all propagators (``all'' mode), 319 master integrals for propagators 1-6 (``loop'' mode), 233 master integrals for propagators 4-6 (``branch'' mode), and the best case, 176 master integrals for the propagator 5 (``propagator'' mode). For topologies where independent internal masses exist, we can do even better. We can simply treat these masses as $\eta$ and thus will not introduce any extra mass scale (``mass'' mode). Because ``mass'' and ``propagator'' mode introduce fewer extra number of master integrals than other modes in general, they usually perform better.

However, as an expense, the boundary analysis is more complicated in general, due to more contributing integration regions as $\eta \to \infty$. Following the general rules of region analysis~\cite{Beneke:1997zp, Smirnov:1999bza}, inequivalent regions can be characterized by the size of loop momentum carried by each branch of the diagram, which can be either of $\mathcal{O}(\sqrt{\eta})$ (denoted as large loop momentum, L) or $\mathcal{O}(1)$ (denoted as small loop momentum, S). For example, for a two-loop integral family with three branches, whose loop momenta can be chosen as $\ell_1$, $\ell_2$ and $\ell_1+\ell_2$, at most five regions may contribute: (LLL), (LLS), (LSL), (SLL) and (SSS).

To obtain boundary conditions, we need to expand the integrands of master integrals in each region. Specifically, in the all-large region (L...L), each propagator should be expanded as
\begin{align}
\frac{1}{((\ell+p)^2-m^2-\kappa\eta)^\nu}\sim\frac{1}{(\ell^2-\kappa\eta)^\nu},
\end{align}
where $\kappa=1$ or $0$, depending on whether $\eta$ is introduced to this propagator or not. We thus obtain vacuum integrals in this region. In the all-small region (S...S), only propagators containing $\eta$ should be expanded as
\begin{align}
\frac{1}{((\ell+p)^2-m^2-\eta)^\nu}\sim\frac{1}{(-\eta)^\nu}.
\end{align}
In this case, we get integrals in a subfamily, with propagators containing $\eta$ contracted. In mixed regions, we need to decompose loop momentum of each propagator as the sum of a large part $\ell_\text{L}$ and a small part $\ell_\text{S}$. Then, if $\ell_L\neq0$ or $\kappa\neq0$, we can expand the propagator as
\begin{align}
\frac{1}{((\ell_\text{L}+\ell_\text{S}+p)^2-m^2-\kappa\eta)^\nu}\sim \frac{1}{(\ell_\text{L}^2-\kappa \eta)^\nu}.
\end{align}
Otherwise, no expansion is needed. After the expansion, the part containing large loop momenta and the part containing small loop momenta are decoupled and we obtain factorized integrals.

\begin{figure}[!h]
\centering
\includegraphics[width=\textwidth]{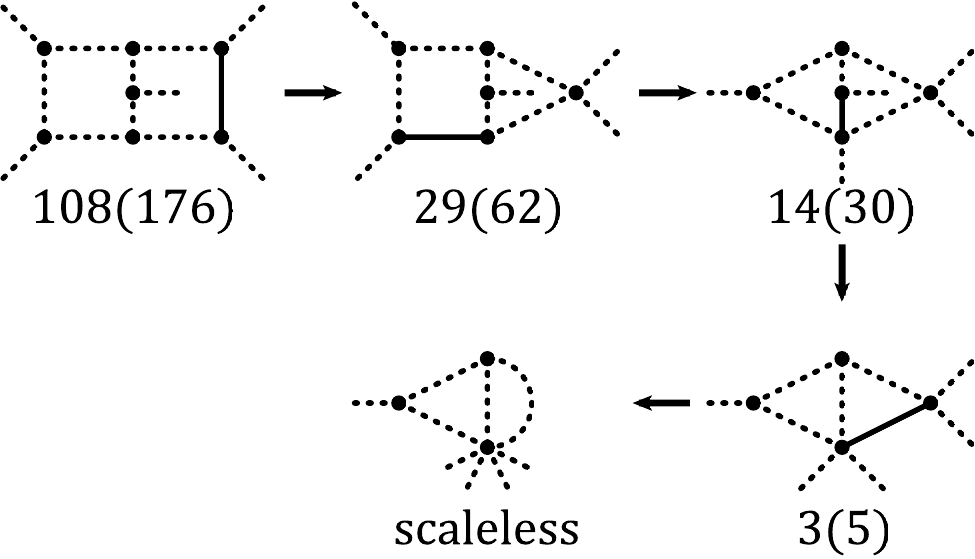}\\
\caption{Figure from Ref.~\cite{Liu:2021wks}. The all-small region iteration of the double-pentagon topology. For each Feynman diagram, the solid line (if exists) represents the propagator where we introduce $\eta$. The number of master integrals of the original topology and the one with $\eta$ introduced (in parentheses) are also listed below each diagram except the last scaleless topology.}\label{fig:dpsim}
\end{figure}

It turns out that usually the boundary integrals are still too complicated to evaluate directly. But this is still fine because they are already simpler than the original integrals, which means we can keep applying the previous procedure to simplify the boundary integrals until they are all known to us. For example, the double-pentagon topology can be simplified iteratively with ``propagator'' mode as shown in Fig.~\ref{fig:dpsim}.

In practice, we would profit from a systematic definition of the terminal topologies. For example, we can always identify single-mass vacuum integrals as our terminals. In Ref.~\cite{Liu:2022mfb}, single-mass vacuum integrals are further simplified in an iterative manner. In that way, we can simply identify the 0-loop integral (whose result is 1) as terminals, which have been proved to be more convenient.


\section{Numerical fit}\label{sec:fit}
A very useful trick implemented in {\tt AMFlow} is numerical fit. Consider a function $f(x)$ which can be expanded near $x=0$ as
\begin{align}\label{eq:fexp}
{f}(x) = \sum_{n=0}^\infty f_n x^n,
\end{align}
and our goal is to compute its estimation up to the $k$-th order
\begin{align}\label{eq:estimation}
\tilde{f}_0+\tilde{f}_1x+\cdots+\tilde{f}_k x^k,
\end{align}
with relative accuracy $E_n\leq E$, where $\tilde{f}_n$ is the estimation of $f_n$ and $E_n$ is defined by
\begin{align}
E_n \equiv \left|\frac{\tilde{f}_n-f_n}{f_n}\right|.
\end{align}

We propose to realize this by evaluating $f(x)$ numerically at some sample points $x_0, x_1, \ldots, x_N$ $(N\geq k)$ near $x=0$ and solving a system of linear equations
\begin{align}\label{eq:linearfit}
\left\{\begin{array}{ccc}
f(x_0)&=&\tilde{f}_0+\tilde{f}_1x_0+\cdots+\tilde{f}_Nx_0^N,\\
f(x_1)&=&\tilde{f}_0+\tilde{f}_1x_1+\cdots+\tilde{f}_Nx_1^N,\\
&\cdots&\\
f(x_N)&=&\tilde{f}_0+\tilde{f}_1x_N+\cdots+\tilde{f}_Nx_N^N.\\
\end{array}\right.
\end{align}
In practice, we find two ways are useful to choose these sample points:
\begin{enumerate}
  \item $|x_0|\sim \cdots\sim|x_N|\sim r \ll R$,
  \item $x_0, \ldots,x_N$ are distributed uniformly on the circle centered at $x=0$ with radius $r<R$,
\end{enumerate}
where $R$ is the convergence radius of the expansion \eqref{eq:fexp}. If one of these ways is chosen and the precision $p$ of the samples $f(x_0),f(x_1),\ldots,f(x_N)$ is sufficiently high, then the relative accuracy of $f_n$ can be roughly estimated as
\begin{align}
E_n \sim\left(\frac{r}{R}\right)^{N+1-n},\quad 0\leq n\leq N.
\end{align}
It can be seen that the relative accuracy $E_n$ decreases as $n$ increases. Thus to achieve our precision goal, we can set $E_k \sim E$, or equivalently
\begin{align}
N \sim k-1+\frac{\log(E)}{\log(r/R)}.
\end{align}
This also gives a constraint about the precision $p$ of the samples
\begin{align}
p \lesssim E_0 \sim \exp \left(k\log(r/R)+\log(E)\right),
\end{align}
because we cannot expect a correct result if the precision of the samples is too low. So the total time consumption to obtain the estimation \eqref{eq:estimation} is
\begin{align}
T &= (N+1)\times t(p)\nonumber\\
&\gtrsim \left(k+\frac{\log(E)}{\log(r/R)}\right)\times t(\exp \left(k\log(r/R)+\log(E)\right)),
\end{align}
where $t(p)$ is the average time needed to compute at a sample point with precision $p$, depending on both the nature of the problem and the numerical algorithm. Typically, in the framework of power series expansion method to solve differential equations of Feynman integrals, the dominant part of $t(p)$ is a polynomial-like object of the number of correct digits, i.e.,
\begin{align}
t(p) \sim (-\log(p))^\alpha,
\end{align}
where $\alpha$ is a positive number. Therefore, we have
\begin{align}
T \gtrsim \left(k+\frac{\log(E)}{\log(r/R)}\right)\times\left(-k\log(r/R)-\log(E)\right)^\alpha,
\end{align}
which can be minimized by choosing
\begin{align}\label{eq:min1}
r\sim R E^{1/(\alpha k)},
\end{align}
and
\begin{align}\label{eq:min2}
N\sim (\alpha+1)k-1, \quad p \lesssim E^{(\alpha+1)/\alpha}.
\end{align}


Next we can discuss how to apply this trick to the computation of Feynman integrals. To
obtain numerical results of master integrals as expansions in $\epsilon$, we can solve differential equations \eqref{eq:deq} with some numerical values of $\epsilon$ and solve a system of linear equations like Eq.~\eqref{eq:linearfit} for each master integral \footnote{Suppose these integrals have been normalized such that they all start with $\epsilon^0$.}. We find the first way to choose sample points stated after Eq.~\eqref{eq:linearfit} is better to use in this case, because we can always choose real values of $\epsilon$ to avoid potential complexities. Note that \eqref{eq:min1} and \eqref{eq:min2} only serve as a reference, and in practice one may need to make some adjustments to get satisfactory results.

This trick brings several benefits. First, a much simpler code structure is made possible, because in this framework all integrals are simply pure numbers rather than expansions in $\epsilon$, which is much easier to carry out. Second, the problem of $\epsilon$-order cancellations is totally resolved, because we never use truncated series in $\epsilon$ to express any integral throughout the calculations, which means we can always include more $\epsilon$-orders by simply increasing the precision of the integrals. Finally, the computations at different sample points are totally independent and thus can be massively parallelized to save our waiting time.

This trick can also be applied to achieve asymptotic expansions of Feynman integrals at a given phase-space point or a given value of $\eta$. Sometimes, this becomes crucial, given the fact that there are usually many removable singularities in differential equations. With the second way to choose sample points on a circle, removable singularities inside the circle can be totally ignored.

\section{Using {\tt AMFlow}}\label{sec:usage}
The latest version of {\tt AMFlow} can be downloaded from
\begin{align}\label{eq:link}
\text{\url{https://gitlab.com/multiloop-pku/amflow}}.
\end{align}
Users can then follow the guidance outlined in \texttt{README.md} to install this package properly on their devices. After that, the package can be loaded by the command\par
\texttt{Get["/path/to/AMFlow.m"];}

\texttt{AMFlow} depends on external programs to do integrals reduction. To use different reducers, one can set the following option
\par
\texttt{SetReductionOptions["IBPReducer" -> {\texttt{\textit{reducer}}}];}\\
where \texttt{\textit{reducer}} can be any reducer whose interface with \texttt{AMFlow} has been built.
Currently, three reducers based on Laporta's algorithm are available, including \texttt{"FiniteFlow+LiteRed"}~\cite{Lee:2013mka, Peraro:2019svx}, \texttt{"FIRE+LiteRed"}~\cite{Lee:2013mka, Smirnov:2019qkx} and \texttt{"Kira"}~\cite{Klappert:2020nbg}. Other reducers can also play their roles after users build their interfaces with {\tt AMFlow} properly.

\begin{figure}[htbp]
\centering
\includegraphics[width=0.4\textwidth]{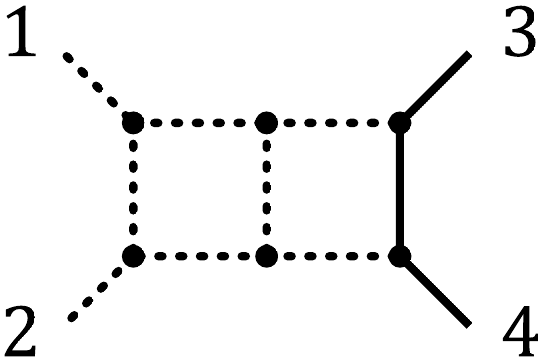}\\
\caption{A two-loop planar integral family involved in NNLO QCD corrections to $t\bar{t}$ hadroproduction.}\label{fig:tt}
\end{figure}

The usage of \texttt{AMFlow} is best illustrated with an example. Fig.~\ref{fig:tt} shows a two-loop planar integral family involved in NNLO QCD corrections to $t\bar{t}$ hadroproduction. The four external momenta $\{p_1, p_2, p_3, p_4\}$ flowing into the diagram satisfy the momentum conservation $p_1+p_2+p_3+p_4=0$ and on-shell conditions $p_1^2=p_2^2 = 0$, $p_3^2=p_4^2=m^2$. Besides, there are two independent kinematic variables $s = (p_1+p_2)^2$ and $t = (p_1+p_3)^2$. The inverse propagators for this diagram can be written as
\begin{align}
\mathcal{D}_1 &= \ell_1^2, \quad \mathcal{D}_2 = (\ell_1+p_1)^2, \quad \mathcal{D}_3 = (\ell_1+p_1+p_2)^2, \quad \mathcal{D}_4 = \ell_2^2,\nonumber\\
\mathcal{D}_5 &= (\ell_2+p_3)^2-m^2, \quad \mathcal{D}_6 = (\ell_2+p_3+p_4)^2,\quad \mathcal{D}_7 = (\ell_1+\ell_2)^2,
\end{align}
and two irreducible scalar products can be chosen as
\begin{align}
\mathcal{D}_8 = (\ell_1-p_3)^2,\quad \mathcal{D}_9 = (\ell_2+p_1)^2.
\end{align}
Suppose our final goal is to compute the following four top-sector integrals
\begin{align}
&I({1,1,1,1,1,1,1,-3,0}), I(1,1,1,1,1,1,1,-2,-1),\nonumber\\
 &I(1,1,1,1,1,1,1,-1,-2), I(1,1,1,1,1,1,1,0,-3),
\end{align}
from $\epsilon^{-4}$ to $\epsilon^0$, with 20-digit coefficients, at a numerical kinematic point
\begin{align}
s = 30,\quad t = -10/3,\quad m^2 = 1.
\end{align}
This example can be found in \texttt{examples/automatic\_vs\_manual/run.wl}.

\subsection{Input}\label{sec:input}
First, we should use the function \texttt{AMFlowInfo} to define globally used objects during the computation, like\par
\texttt{AMFlowInfo[\textit{key}] = \textit{obj};}\\
where \texttt{\textit{key}} should be a string pre-defined in \texttt{AMFlow} and \texttt{\textit{obj}} should be the corresponding object. We list most frequently used pre-defined strings and the meaning of their corresponding objects below:\par
\texttt{"Family"} - the name of the integral family;\par
\texttt{"Loop"} - a list of all loop momenta;\par
\texttt{"Leg"} - a list of all external momenta;\par
\texttt{"Conservation"} - a list of replacement rules for momentum conservation;\par
\texttt{"Replacement"} - a list of complete replacement rules for scalar products among external legs;\par
\texttt{"Propagator"} - a list of complete inverse propagators;\par
\texttt{"Numeric"} - a list of replacement rules indicating the numerical kinematics where to perform the computation;\par
\texttt{"NThread"} - the number of threads in use.\\
In this example, we can simply write, e.g.,\par
\texttt{AMFlowInfo["Family"] = tt;}\par
\texttt{AMFlowInfo["Loop"] = \{l1, l2\};}\par
\texttt{AMFlowInfo["Leg"] = \{p1, p2, p3, p4\};}\par
\texttt{AMFlowInfo["Conservation"] = \{p4 -> -p1-p2-p3\};}\par
\texttt{AMFlowInfo["Replacement"] = \{p1\^{}2 -> 0, p2\^{}2 -> 0, \\ p3\^{}2 -> msq,
p4\^{}2 -> msq, (p1+p2)\^{}2 -> s, (p1+p3)\^{}2 -> t\};}\par
\texttt{AMFlowInfo["Propagator"] = \{l1\^{}2, (l1+p1)\^{}2, (l1+p1+p2)\^{}2,  \\l2\^{}2,
(l2+p3)\^{}2-msq, (l2+p3+p4)\^{}2, (l1+l2)\^{}2, (l1-p3)\^{}2, \\(l2+p1)\^{}2\};}\par
\texttt{AMFlowInfo["Numeric"] = \{s -> 30, t -> -10/3, msq -> 1\};}\par
\texttt{AMFlowInfo["NThread"] = 4;}\\

\subsection{Automatic computation}\label{sec:auto}
\texttt{AMFlow} provides a function named \texttt{SolveIntegrals} to perform automatic computations of Feynman integrals.
A general usage of this function should be like\par
\texttt{auto = SolveIntegrals[target, goal, epsorder];}\\
where \texttt{target} is a list of target integrals, \texttt{goal} represents the precision goal and \texttt{epsorder} means the length of $\epsilon$ expansion in the final expansions, i.e., starting from $\epsilon^{-2L}$ and ending at $\epsilon^{-2L+\text{\texttt{order}}}$ with $L$ the number of loops. This function will first reduce the target integrals to master integrals and then compute master integrals using auxiliary mass flow. The output \texttt{auto} is a list of replacement rules from integrals to their values.

For current example, we can write\par
\texttt{target=\{j[tt,1,1,1,1,1,1,1,-3,0],j[tt,1,1,1,1,1,1,1,-2,-1],}\par
\texttt{\;\;\;\;\;\;\;\;\;\;j[tt,1,1,1,1,1,1,1,-1,-2],j[tt,1,1,1,1,1,1,1,0,-3]\};}\par
\texttt{goal=20;}\par
\texttt{epsorder=4;}\\
where we have adopted the notation of \texttt{LiteRed} to represent an integral
\begin{align}
\text{\texttt{j[tt,$\nu_1$,...,$\nu_9$]}} \quad\leftrightarrow\quad I(\nu_1,\ldots,\nu_9) \text{ in family \texttt{tt}}.
\end{align}
After the computation, the output \texttt{auto} should be like (if \texttt{"FiniteFlow+LiteRed"} is chosen)\par
\texttt{\{j[tt,1,1,1,1,1,1,1,-2,-1] -> }\par
\texttt{-0.029131054131054131054/eps\^{}4}\par
\texttt{+0.15634543151250003740/eps\^{}3}\par
\texttt{-(0.007823397125433531023-0.138772644276800647440I)/eps\^{}2}\par
\texttt{+(6.4018478848121593013-5.3096594693278082225I)/eps}\par
\texttt{+(6.385202185942958097+49.103186001778095122I),}\par
\texttt{...\}}\\
where \texttt{eps} means the dimensional regulator $\epsilon$.

\subsection{Manual computation}\label{sec:man}

Although \texttt{SolveIntegrals} is designed for most general purposes, there could be some extreme cases where this function may not be able to produce satisfactory results. So we introduce a more involved way to compute integrals in this section.

We first use the function \texttt{GenerateNumericalConfig} to regenerate the parameters for numerical evaluation suggested by \texttt{SolveIntegrals}\par
\texttt{\{epslist, workingpre, xorder\} = GenerateNumericalConfig[}\par
\texttt{goal, epsorder];}\\
where \texttt{goal} and \texttt{epsorder} have been defined in the previous section. The output is a triblet: \texttt{epslist} is a list of suggested sample points of $\epsilon$, \texttt{workingpre} is the suggested working precision and \texttt{xorder} is the suggested truncated order of the power series expansions. In principle, if these suggested parameters are used, we will obtain exactly the same results as \texttt{SolveIntegrals}. So, when \texttt{SolveIntegrals} fails to generate satisfactory results, users can define their own
\texttt{epslist}, \texttt{workingpre} and \texttt{xorder}.

We then tell the program our preferred parameters by\par
\texttt{SetAMFOptions["WorkingPre"->workingpre, "XOrder"->xorder];}\\
and compute target integrals on $\epsilon$-samples by\par
\texttt{soleps = BlackBoxAMFlow[target, epslist];}\\
where \texttt{target} is the list of target integrals defined in the previous section. The output \texttt{soleps} is a list of replacement rules like\par
\texttt{\{j[tt,1,1,1,1,1,1,1,-2,-1] -> \{v11,v12,...,v1n\}, }\par
\texttt{j[tt,1,1,1,1,1,1,1,-1,-2] -> \{v21,v22,...,v2n\}, ...\}}\\
where \texttt{vij} are pure numbers, representing the value of the $i$-th integral on the $j$-th $\epsilon$-sample. After that, we need to fit the expansions in $\epsilon$ using these samples. This can be achieved by\par
\texttt{exp = FitEps[epslist, \#, leading]\&/@Values[soleps];}\\
where \texttt{leading} means the leading power of $\epsilon$-pole in the expansions, which can be set to $-4$ in this example. The output \texttt{exp} is just the list of expansions in \texttt{eps} for target integrals.

\subsection{Other functions}\label{sec:other}
There are also other useful functions in \texttt{AMFlow}. Here we just give a brief summary. For more details, users can investigate corresponding examples provided in the folder \texttt{examples}.
\begin{enumerate}
  \item Computation of integrals containing linear propagators~\cite{Liu:2022tji}.\\
        See \texttt{linear\_propagator}.
  \item Computation of phase space integrations~\cite{Liu:2020kpc}.\\
        See \texttt{aotumatic\_phasespace} and \texttt{feynman\_prescription}.
  \item Computation of asymptotic expansions using the differential equations solver provided in \texttt{AMFlow}, either by traditional matching or numerical fit introduced in section~\ref{sec:fit}.\\
        See \texttt{differential\_equation\_solver}.
  \item Computation of integrals with complex kinematic parameters. \\
        See \texttt{complex\_kinematics}.
  \item Computation of integrals in arbitrary space-time dimension. \\
        See \texttt{spacetime\_dimension}.
\end{enumerate}

\subsection{Summary of options}

\texttt{AMFlow} allows users to set global options through \texttt{SetAMFOptions}, \\ \texttt{SetReductionOptions} and \texttt{SetReducerOptions}. Here we list and describe the most frequently used options. For other options, we refer the users to the file \texttt{options\_summary}.

\begin{longtable}[htp]{|p{4.3cm}|p{8.5cm}|}
\hline
Option and Default & Description \\ \hline
\multicolumn{2}{|c|}{\texttt{SetAMFOptions}} \\ \hline
\texttt{"D0"}$\to$\texttt{4} &  A rational number $D_0$ such that the integrals will be computed with $D = D_0-2\epsilon$.\\\hline
\texttt{"WorkingPre"}$\to$\texttt{100} & Working precision when performing numerical computations, including solving differential equations and fitting.\\\hline
\texttt{"XOrder"}$\to$\texttt{100} & Truncated order of expansions when solving differential equations.\\\hline
\multicolumn{2}{|c|}{\texttt{SetReductionOptions}}  \\ \hline
\texttt{"IBPReducer"}$\to$ \texttt{"FiniteFlow+LiteRed"} & Integration-by-parts reducer. Available reducers include \texttt{"FiniteFlow+LiteRed"}, \texttt{"Kira"} and \texttt{"FIRE+LiteRed"}.\\\hline
\texttt{"BlackBoxRank"}$\to$\texttt{3} & Suggested maximal rank of seed integrals when constructing IBP systems. But if the maximal rank of target integrals $s$ is larger than the value of this option, then the maximal rank of seed integrals will be adjusted to $s$ internally.\\\hline
\texttt{"BlackBoxDot"}$\to$\texttt{0} & Suggested maximal dot of seed integrals when constructing IBP systems. But if the maximal dot of target integrals $r$ is larger than the value of this option, then the maximal dot of seed integrals will be adjusted to $r$ internally.  \\\hline
\multicolumn{2}{|c|}{\texttt{SetReducerOptions} (\texttt{"FiniteFlow+LiteRed"} or \texttt{"FIRE+LiteRed"} used)}  \\\hline
\texttt{"EMSymmetry"}$\to$\texttt{False} & A parameter indicating whether symmetries among external legs should be exploited when preparing the topology using \texttt{LiteRed}.\\\hline
\multicolumn{2}{|c|}{\texttt{SetReducerOptions} (\texttt{"Kira"} used)}  \\\hline
\texttt{"IntegralOrder"}$\to$\texttt{5} &  A positive integer ranging from 1 to 8 specifying the integral ordering for \texttt{Kira}. For more details, see Ref.~\cite{Maierhofer:2018gpa}.\\\hline
\texttt{"ReductionMode"}$\to$ \texttt{"Kira"} & Reduction mode for \texttt{Kira}. Available modes include \texttt{"Kira"}, \texttt{"FireFly"}, \texttt{"Mixed"} and \texttt{"NoFactorScan"}. See Ref.~\cite{Klappert:2020nbg} for more details.\\\hline
\end{longtable}

\section{Summary and outlook}\label{sec:conc}

In this paper, the Mathematica package \texttt{AMFlow} is presented together with some explicit examples. We have highlighted the numerical fit strategy, which can overcome many difficulties when numerically solving differential equations.

The differential equations solver provided in \texttt{AMFlow} is of high performance and very suitable for high precision computations. In later version of {\tt AMFlow}, we will provide some functions for users to access this solver in a more convenient way.

With the auxiliary mass flow method, integral reduction will be the only input for calculating Feynman integrals~\cite{Liu:2022mfb}. In the near future, a public implementation of the reduction method developed in Refs.~\cite{Liu:2018dmc, Guan:2019bcx} will be available, which can typically reduce the time consumption by 2 orders of magnitude comparing with other methods on the market. With this powerful reduction package and \texttt{AMFlow}, complicated integrals such as those in Ref.~\cite{Liu:2021wks}, can be computed automatically.

\section*{Acknowledgements}
We thank Z. F. Liu and C. Y. Wang for many useful discussions. This work was supported in part by the National Natural Science Foundation of China (Grants No. 11875071, No. 11975029), the National Key Research and Development Program of China under Contracts No. 2020YFA0406400 and the High-performance Computing Platform of Peking University. The research of XL was also supported by the ERC Starting Grant 804394 \textsc{HipQCD} and by the UK Science and Technology Facilities Council (STFC) under grant ST/T000864/1.

\bibliographystyle{utphysMa}
\providecommand{\href}[2]{#2}\begingroup\raggedright\endgroup


\end{document}